\documentclass[aps,pre,amsmath,amssymb,lengthcheck,showpacs,superscriptaddress]{revtex4}
\usepackage{epsfig}
\usepackage{graphicx}
\usepackage{dcolumn}
\usepackage{bm}
\usepackage[english]{babel}

\begin{document}

\title{Extension of Classical Nucleation Theory for Uniformly Sheared Systems}

\author{
Anatolii V. Mokshin } \email{anatolii.mokshin@mail.ru}
\affiliation{Kazan Federal University, 420000 Kazan,  Russia }

\author{
Bulat N. Galimzyanov} \email{bulatgnmail@gmail.com}
\affiliation{Kazan Federal University, 420000 Kazan,  Russia }

\author{
Jean-Louis Barrat} \email{jean-louis.barrat@ujf-grenoble.fr}
\affiliation{Laboratory for Interdisciplinary Physics, UMR 5588,
Universit\'{e} Grenoble 1 and CNRS, 38402 Saint Martin
d'H\`{e}res, France }

%

\date{\today}

\begin{abstract}
Nucleation is an out-of-equilibrium process, which can  be
strongly affected by the presence of external fields. In this
letter, we report a simple extension of classical nucleation
theory to systems submitted to an homogeneous shear flow.  The
theory involves accounting for the anisotropy of the critical
nucleus formation, and introduces a shear rate dependent effective
temperature.  This extended theory is used to analyze the results
of extensive molecular dynamics simulations, which explore a broad
range of shear rates and undercoolings.  At fixed temperature, a
maximum in the nucleation rate is observed, when the relaxation
time of the system is comparable to the inverse shear rate. In
contrast to previous studies, our approach does not require a
modification of the thermodynamic description, as the effect of
shear is mainly embodied into a modification of the kinetic
prefactor and of the temperature.
\end{abstract}


\maketitle

\section{Introduction \label{intr}}
Homogeneous crystal nucleation is the route by which the
crystallization in a supercooled liquid is initiated, at least in
the absence of impurities or of a spinodal
instability~\cite{Kelton_book_1991}. A general physical
understanding this phenomena is provided by the classical nucleation
theory (CNT), which takes into account the thermodynamic and kinetic
aspects properly to treat the formation of the crystalline nuclei
that are able to
grow~\cite{Kelton_book_1991,Kashchiev_Nucleation_2000,Frenkel_book_1946}.
Nevertheless, despite  the strong theoretical background of the CNT,
there are still debates on some fundamental questions. One example
is the issue of the mechanisms, -- spinodal decomposition or
nucleation -- by which the structural ordering is started in a
simple one-component system at extremely deep
supercooling~\cite{Parrinello_PRL_2006}.

Imposing external driving fields (e.g. a shear flow) leads to
nonequilibrium steady state of the system and, thereby, impacts on
the nucleation process in a complex manner. This is verified by
several studies of the shear-induced effects on the structural
ordering in colloidal
suspensions~\cite{Butler/Harrowell_PRE_1995,Pusey_PRE_1998,Cerda_PRE_78_2008,Shereda_PRL_228302_2010,Blaak_PRL_93_2004},
polymers~\cite{Olmsted_PRL_103_2009}, $2$D dusty
plasma~\cite{Morfill_PRL_108_2012} and
glasses~\cite{Mokshin/Barrat_PRE_2008}. Although  kinetic models
for nonequilibrium nucleation have been
proposed~\cite{Onuki_JPCM_1997}, the possibility of a simple
extension of the CNT to nonequilibrium situations is still
open~\cite{Lowen_review_2008}. Moreover, it was recently found
that the mechanism of the shear-induced structural ordering
depends also on the character of the applied shear. According to
results of Ref.~\cite{Shereda_PRL_228302_2010}, the nucleation
mechanism, which is realized in colloidal suspensions under
homogenous shear field, is changed  into crystallization through
the `front propagation' scenario for inhomogeneous, wall-driven
shear. Interestingly, such an ordering  mechanism was not observed
for the model glassy systems under \textit{inhomogeneous
wall-driven shear
flow}~\cite{Mokshin/Barrat_2009_2010,Mousseau_PRB_2011}, where the
nucleation events were clearly detected. Nevertheless, the lack of
a comprehensive study of the impact of the \textit{homogeneous
shear} on the structural ordering in glassy materials (especially,
at deep supercooling) motivates  further investigations.

In this paper, we study the influence of an \emph{homogeneous} shear
drive on the crystal nucleation in a \textit{bulk glassy system}
on the basis of the nonequilibrium molecular dynamics. We analyze
in detail the morphology of the ordered structures and the
statistics of the nucleation events, that allows us to obtain
independently all the ingredients that enter the nucleation rate
and to evaluate directly the mechanisms of the ordering. The
nucleation rates obtained at several temperatures and values of
the shear rate $\dot{\gamma}$ indicate that the steady shear
deformation can either enhance or suppress crystal nucleation,
depending on the magnitude of $\dot{\gamma}$. They can be
described quantitatively using an extension of CNT involving an
effective, shear rate dependent temperature.

\section{Simulation details \label{model_system}}
Molecular dynamics simulations are performed for a single-component
glass forming system,  made of  particles interacting via a
short-ranged oscillatory potential, suggested originally by
Dzugutov~\cite{Dzugutov_MCT}~\footnote{The terms $\epsilon$ and
$\sigma$ define the units of energy and length, respectively. Time,
pressure and temperature units are measured in $\tau = \sigma
\sqrt{m_0/\epsilon}$, $\epsilon/\sigma^{3}$ and $\epsilon/k_{B}$,
respectively.}. At zero pressure, this system is characterized by a
melting temperature $T_m \simeq 1.02 \epsilon/k_B$, and  a glass
transition  at $T_c < T_{MCT} = 0.4 \epsilon/k_B$
\cite{Dzugutov_MCT}.  An applied pressure $P=14\epsilon/\sigma^3$
shifts the system melting and glass temperatures to the values $T_m
\simeq 1.51 \epsilon/k_B$ and $T_c \simeq 0.65 \pm 0.1
\epsilon/k_B$, respectively~\footnote{These temperatures have been
determined from an accurate estimate of the potential energy and
density variation as well as translational and orientational order
parameters at several $(P,T)$-points. The glass transition is
verified by evaluating the particle pair distributions and mean
square displacements.}. Simulation cells of fixed volume $V=\ell^3$
($\ell=20.03\sigma$) containing an amorphous sample with periodic
boundary conditions were prepared by quenching from a well
equilibrated melt at the temperature $T=2.3\epsilon/k_B$ to the
temperatures $T=0.05$, $0.1$, $0.15$, $0.3$ and $0.5\;\epsilon/k_B$,
below $T_c$, with the  quenching rate $0.001\;\epsilon/(k_B \tau)$.
At each temperature, a set of hundred independent configurations
(each is consisting of $6\;912$ particles) was prepared in view of a
statistical analysis.

After the quench, an  homogeneous shear flow is imposed by means
of the SLLOD algorithm supplemented by Lees-Edwards periodic
boundary conditions~\cite{Frenkel_book_2001}. Here, the $x$-axis
is associated with shear direction, the $y$-axis corresponds to
gradient direction and the $z$-axis coincides with vorticity
direction, and the shear rate has a constant value $\dot{\gamma}$
throughout an each run. Constant temperature and pressure
conditions are ensured by using the Nos\'{e}-Hoover method with an
external  pressure $P=14\epsilon/\sigma^3$ that promotes
crystallization in the system. The homogeneous character of the
shear flow is verified from the linearity of velocity profiles.

\section{Results \label{results}}
\subsection{Mechanism of ordering and crystalline structures}
To identify the particles involved in the crystalline phase, the
environment  within the first coordination shell of each particle is
analyzed in terms  of the bond orientational order parameters
\cite{Steinhardt_PRB_1983}, and the corresponding clusters are
constructed. The time-dependent cluster size distribution
$\mathcal{N}_n(t)$ is evaluated for each run, and is averaged over
independent runs. Before going to the quantitative analysis of these
data, we describe briefly some qualitative aspects of the nucleation
process.
\begin{figure}[tbh] 
\centerline{\psfig{figure=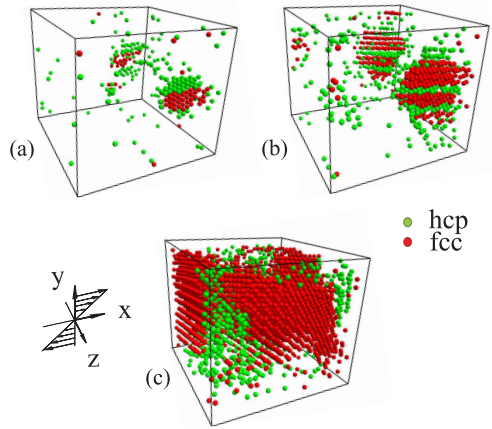,height=6.3cm,angle=0}}
\caption{(color online) Configurations of the particles belonging to
the crystalline phase for a sample sheared with the rate
$\dot{\gamma}=0.001\tau^{-1}$ at a temperature $T=0.1\epsilon/k_B$
and at different times: (a) $70\tau$, (b) $110\tau$ and (c)
$180\tau$. Colors are associated with the types of crystalline
arrangement (hcp and fcc) as defined from the cluster analysis.
Results for other temperatures and shear rates are similar to those
presented. \label{fig_snapshots}} 
\end{figure}

The cluster data analysis reveals the following features as
structural order appears  in the system. The nuclei of a fcc/hcp
crystal phase are homogeneously distributed, and their sizes
fluctuate while they are lower than some critical size. When
reaching a critical size, they start growing monotonously with
time. These are clear indications that a homogeneous nucleation
mechanism is at work.  Similar observations are made at all shear
rates $\dot{\gamma} \in [0,\;0.01]\; \tau^{-1}$ and temperatures,
even at a deep supercooling $(T_m-T)/T_m \simeq 0.97$.
Figure~\ref{fig_snapshots} shows, as an example, the
configurations of particles generating the crystalline phases for
a single amorphous sample at the temperature $T=0.1\;\epsilon/k_B$
for a shear rate $\dot{\gamma}=0.001\;\tau^{-1}$ at different
times after startup of the steady shear. It is seen that this type
of  shear-driven structural ordering differs completely from the
results of Ref.~\cite{Shereda_PRL_228302_2010} reported recently
for shear-induced colloidal crystallization, where shear was
applied through the walls and ordering appears as propagated layer
by layer. Instead of the crystallization through a layering, a
nucleation-growth mechanism is clearly observed here,  similar to
the results of Ref.~\cite{Blaak_PRL_93_2004} for a colloidal
system and of Ref.~\cite{Olmsted_PRL_103_2009} for flow-induced
ordering in polymers.

\begin{figure}[tbh] 
\centerline{ \psfig{figure=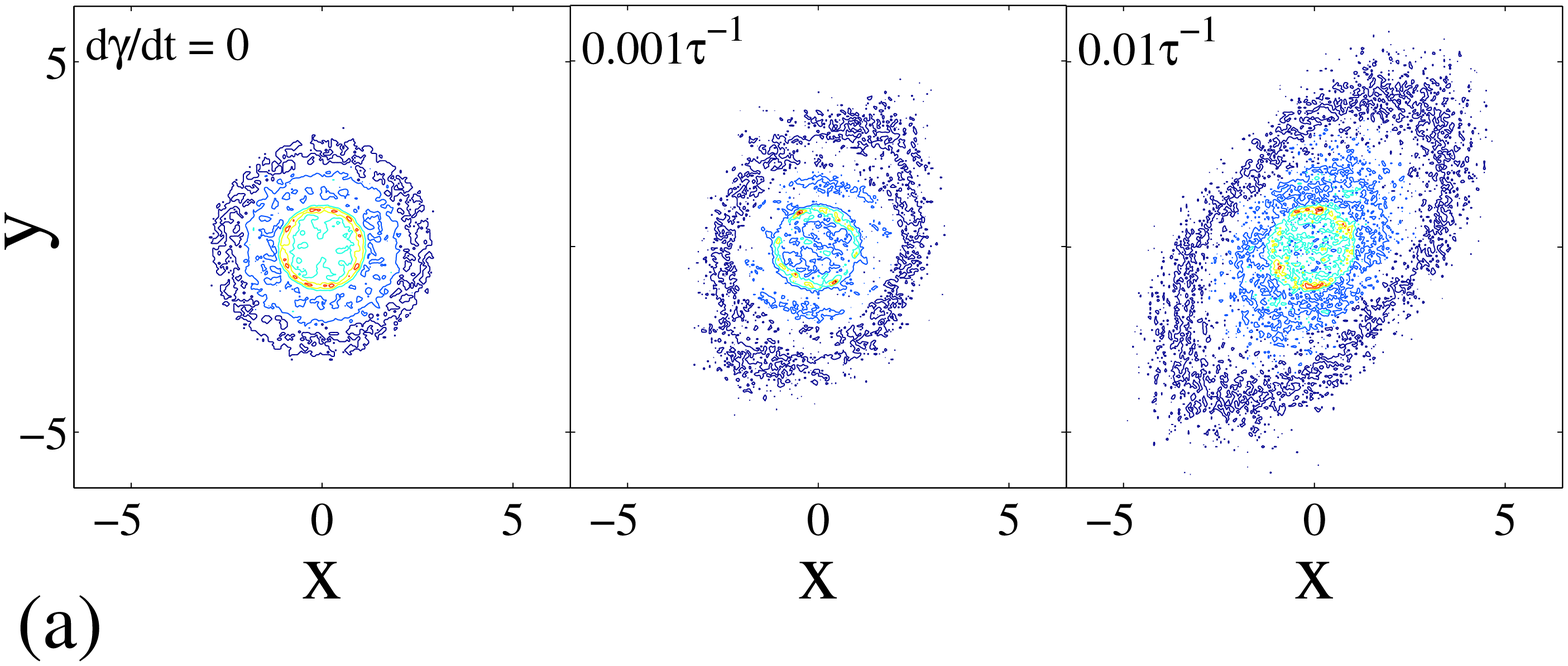,height=4.5cm,angle=0}}
\centerline{\psfig{figure=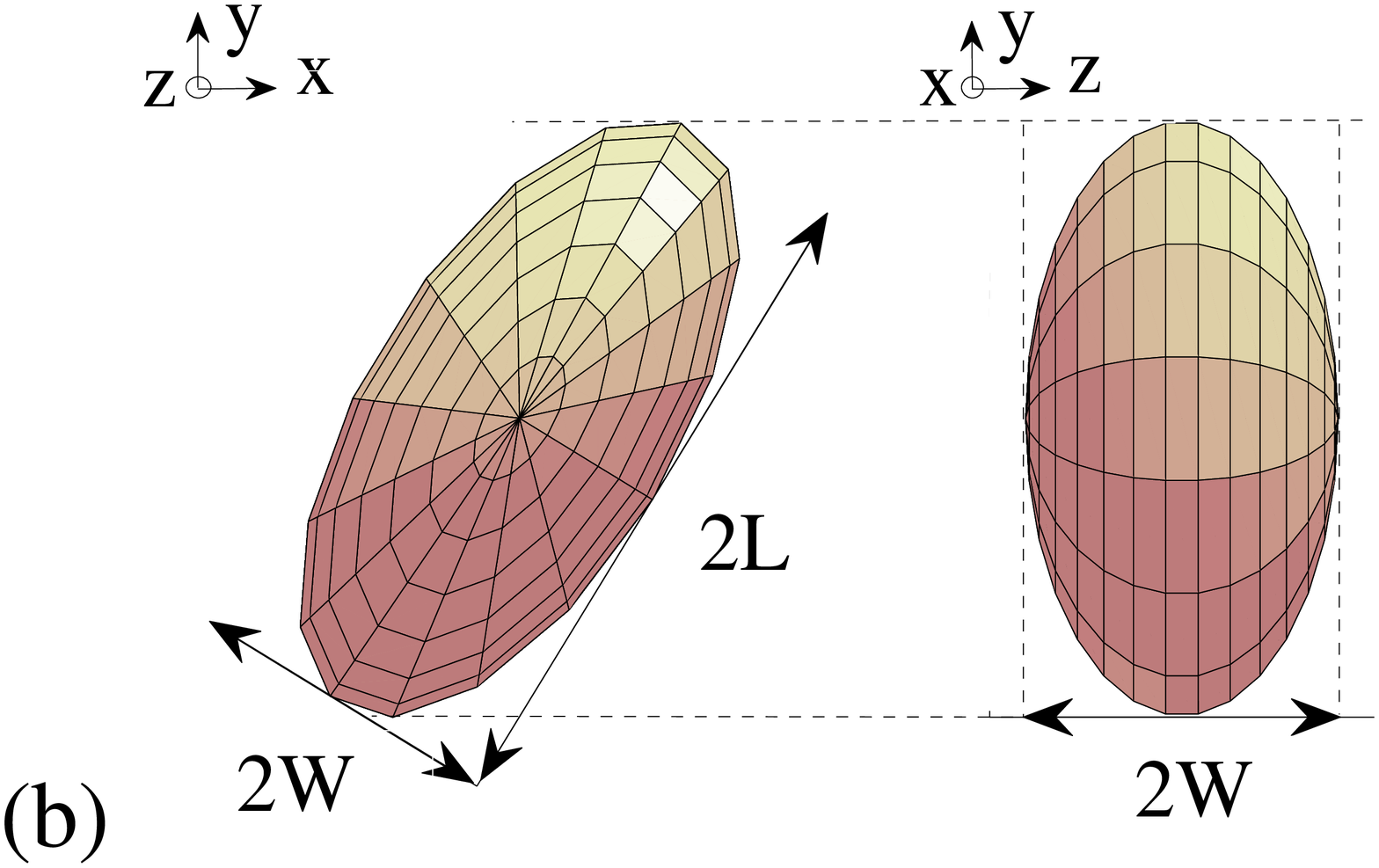,height=3.2cm,angle=0}}
\caption{(color online) (a) Correlation function $g(x,y)$ of the
nucleated cluster for a temperature $T=0.15\epsilon/k_B$ in the
zero-shear case as well as for finite shear rates. Coordinates are
given in units of $\sigma$. Shear-flow produces a critical nucleus
of ellipsoid shape and tilts  its  symmetry axis within
$xy-$plane. (b) Schematic drawing of the envelope of the critical
cluster, which takes an ellipsoidal shape oriented in the
$xy$-plane.\label{fig_pair_corr_fucn}} 
\end{figure}

A key ingredient in the analysis of the shear-driven nucleation will
be  the geometry of the nucleated clusters.  To quantify this
geometry, we define the pair correlation functions, $g(x,y)$ and
$g(x,z)$, which are computed only for the particles involved in a
critical cluster and characterize the distribution of the particles
projected onto $xy-$ and $xz-$ planes, respectively. Our results
show that  the critical nucleus changes from a spherical shape  in
the  shear-free case to a prolate ellipsoid with its long axis
tilted in the $xy-$plane (see Fig.~\ref{fig_pair_corr_fucn}).
Numerically, the ellipticity $\varepsilon=W/L$ grows with increasing
shear rates. For the highest shear rate $\dot{\gamma}=0.01\tau^{-1}$
it takes $\varepsilon=0.83$ at $T=0.05\epsilon/k_B$ and
$\varepsilon=0.75$ at $T=0.5\epsilon/k_B$. Such flow influence on a
shape of the nuclei corresponds to the recent results of Graham and
Olmsted obtained within coarse-grained simulations of flow in
polymer melts~\cite{Olmsted_PRL_103_2009}.

\subsection{Nucleation under shear drive}
We now turn to the quantitative analysis of the nucleation data,
which is based on the study of the time dependent cluster
distribution $\mathcal{N}_n(t)$. In the framework of a mean first
passage analysis~\cite{Bartell_Wedekind_MFPT}, the average time
$\tau(n)$, at which a cluster of a given size $n$ is observed for
the first time, is analyzed together with its derivative $\partial
\tau(n)/\partial n$~\cite{Mokshin/Galimzyanov_JPCB_2012}. This
allows one to obtain independently the following  quantities
characterizing the nucleation process: the steady-state nucleation
rate $J_{s}=1/( V \tau_c)$ ($\tau_c$ is the nucleation time scale),
the number $n_c$ of particles involved in the critical nucleus, and
the Zeldovich factor $Z$ which characterizes the curvature of the
free energy barrier at the top
\cite{Mokshin/Barrat_PRE_2008,Mokshin/Barrat_2009_2010}. In
addition, the interfacial free energy $\gamma_m$ was estimated by
means of the thermodynamic integration of the surface energy of the
critical nuclei~\cite{Mokshin/Galimzyanov_JPCB_2012}. The different
nucleation parameters are related by the general expression of the
nucleation rate within the
CNT~\cite{Kashchiev_Nucleation_2000,Frenkel_book_1946}:
\begin{equation} \label{eq: nucleation_rate}
J_{s} = g_{n_c}^{+}Z \rho_{am} \exp \left ( - \Delta G_{n_c} / k_B
T  \right ).
\end{equation}
Here, $g_{n_c}^{+}$ is the rate at which atoms attach to the
critical nucleus, $\rho_{am}$ is the density of atoms in the
parent amorphous phase, $\Delta G_{n_c}$ is the free energy
required to form a cluster of the critical size $n_c$, that
corresponds to the maximum in the free energy
\begin{equation}
\Delta G(n) = \Delta G_{n_c} \left[ 3\left( \frac{n}{n_c}
\right)^{2/3}  -2 \left( \frac{n}{n_c} \right) \right],
\end{equation}
and the Zeldovich factor is defined as
\begin{eqnarray} \label{eq: Zeldovich_factor}
Z^2 &=&  \frac{-1}{2 \pi k_B T} \left [ \frac{\partial^2 \Delta
G(n)}{\partial n^2} \right ]_{n=n_c} = \frac{1}{3\pi n_c^2}
\frac{\Delta G_{n_c}}{k_B
T} \nonumber \\
&=& \frac{\gamma_m}{9 \pi k_B T} \left ( \frac{36 \pi}{ \rho_{c}^2
n_c^4} \right )^{1/3}.
\end{eqnarray}
The last equality in Eq.~(\ref{eq: Zeldovich_factor}) results
from the assumption that the nascent clusters have a spherical
form~\cite{Kashchiev_Nucleation_2000}.
\begin{figure}[tbh]  
\centerline{ \psfig{figure=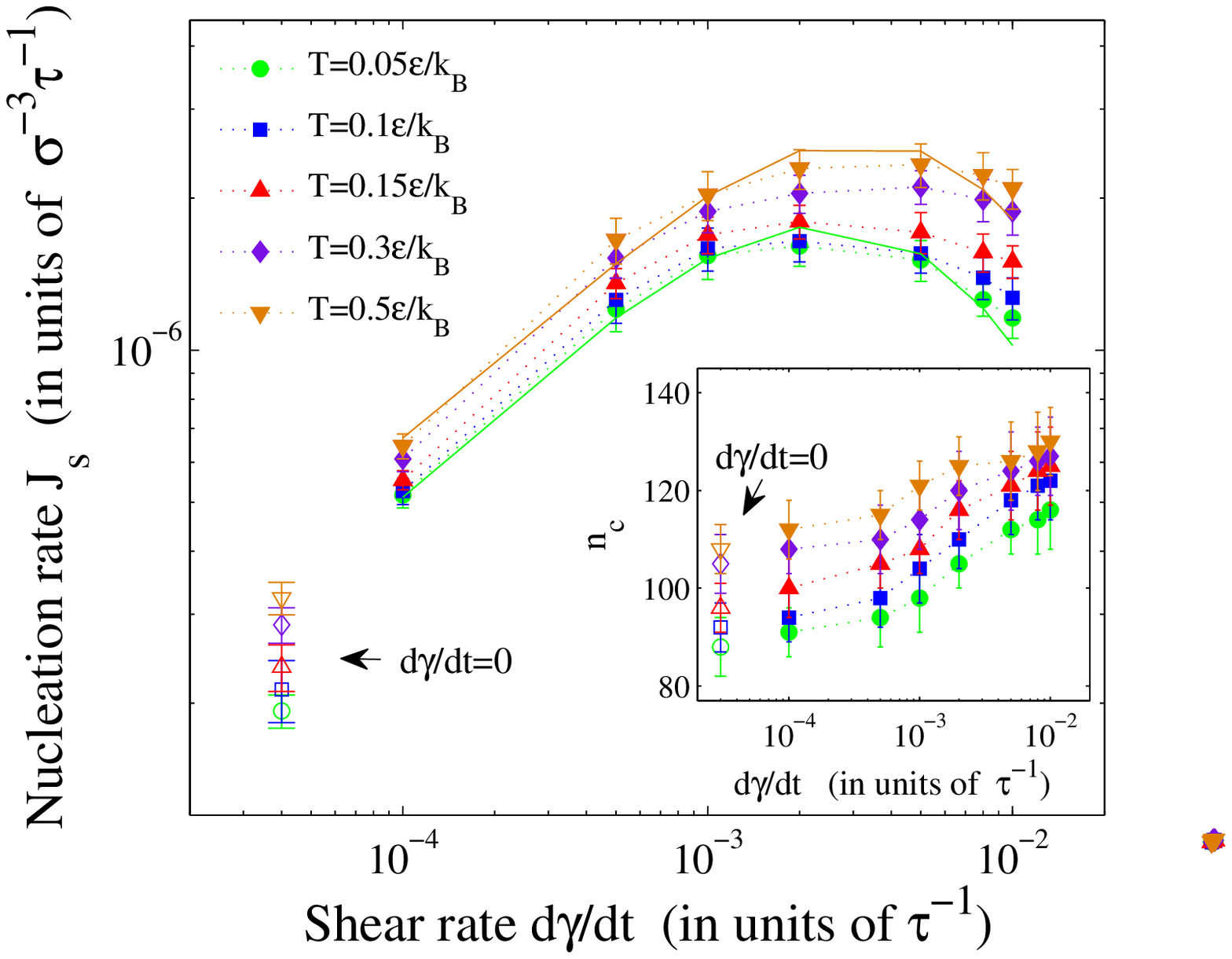,height=6.7cm,angle=0}}
\caption{ (color online) Main: Nucleation rate vs shear rate at
 five different  temperatures. The dashed line is the fit
by $J_s(\dot{\gamma},T) = J_s(\dot{\gamma}=0,T) + (A
\dot{\gamma}/V)
\exp[-(\alpha(T)^2\dot{\gamma}\tau_{\alpha}(T))^n]$, where $A =
33$ and $n=1/2$ are the temperature independent dimensionless
parameters, $\tau_{\alpha}(T)$ is the structural relaxation time
defined for the shear-free cases, and $\alpha(T)$ changes from
$4.5$ at $T=0.05\epsilon/k_B$ to $9.0$ at $T=0.5\epsilon/k_B$.
Inset: Critical cluster size vs shear rate at the different
temperatures. \label{fig_nucl_rate}} 
\end{figure}

To interpret our results we extend the CNT above in two
directions: (i) the calculation of the nucleation barrier is
extended to ellipsoidal nuclei, and (ii) the temperature is
considered as a free parameter, $T_{\mathrm{eff}}$, in the spirit
of the effective temperature concept~\cite{Cugliandolo_JPA_2011}.
The shape of the nuclei is directly obtained from the simulations,
and is not an adjustable parameter. Therefore, the
overdetermination indicated above allows one to find independently
the two unknown quantities, $T_{\mathrm{eff}}$ and  $g_{n_c}^{+}$,
both of which are expected to be $\dot{\gamma}$-dependent. Thus,
we have
\begin{equation} \label{eq: nucleation_rate_eff}
J_{s} = g_{n_c}^{+}Z \rho_{am} \exp \left ( - \Delta G_{n_c} / k_B
T_{\mathrm{eff}} \right ),
\end{equation}
where $Z$ for a prolate spheroid nucleus of density $\rho_c$ with
ellipticity $\varepsilon$ takes the form
\begin{equation} \label{eq: Zeldovich_factor_eff}
Z =   \frac{1}{n_c^{2/3}} \left ( \frac{\alpha_n}{3 \pi k_B
T_{\mathrm{eff}}} \right )^{1/2}.
\end{equation}
The term $\alpha_n$ is related with the surface contribution into
the nucleation barrier $\Delta G_{n_c}$ and can be written as
\begin{equation}
\alpha_n = \gamma_m \left (  \frac{\pi \varepsilon^2}{6 \rho_c^2}
\right )^{1/3} \left [1 +
\frac{\arcsin{(\sqrt{1-\varepsilon^2})}}{\varepsilon
\sqrt{1-\varepsilon^2}} \right ]. \label{eq: surf_term}
\end{equation}
For the case of spherical nuclei, $\varepsilon \rightarrow 1$,
Eqs.~(\ref{eq: Zeldovich_factor_eff}) and (\ref{eq: surf_term})
yield the last equality of Eq.~(\ref{eq: Zeldovich_factor}).

\subsection{Nucleation rate and critical size}
In Fig.~\ref{fig_nucl_rate}, we show the extracted values of the
nucleation rate $J_s$ and the critical cluster size $n_c$ at
different shear rates $\dot{\gamma}$ and temperatures~$T$. In the
shear-free limit, $n_c$ increases with temperature as expected from
thermodynamics. The term $J_s$,  on the other hand, is a decreasing
function of $T$, which  indicates that the slowing down due to the
viscosity dominates the temperature evolution.  It may seem
surprising to observe nucleation in very low temperature systems, in
which the viscosity is effectively infinite. However, studies of
aging indicate that immediately after the quench the relaxation time
is finite \cite{KobBarrat2000}, so that the local structural
rearrangements that lead to nucleation are possible.
\begin{figure}[tbh] 
\centerline{ \psfig{figure=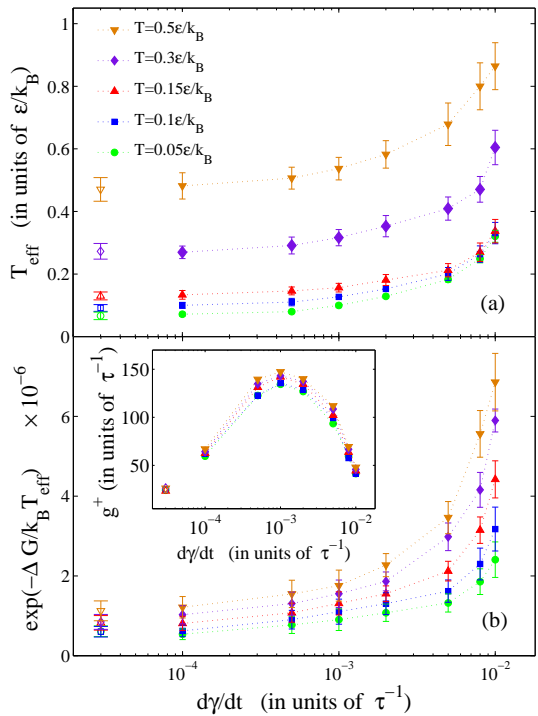,height=10.0cm,angle=0}}
\caption{(color online) (a) Effective temperature vs shear rate at
tdifferent temperatures. (b) Main: Shear rate dependence of the
Arrhenius exponential at different temperatures. Inset: Shear rate
dependence of the kinetic prefactor $g^+_{n_c}$. The open symbols
on the left indicate the values at $\dot{\gamma}=0$.
\label{fig_eff_temp}} 
\end{figure}

At low shear rates, the nucleation rate increases essentially
linearly with $\dot{\gamma}$, as indicated by the empirical
fitting formula given in Fig.~\ref{fig_nucl_rate}. In this regime,
the size $n_c$ appears to be practically unchanged and a shape
remains spherical. Consequently, the thermodynamic contribution
can be considered as unaffected by shear-flow and the nucleation
is enhanced, mainly, through the kinetic contribution $g_{n_c}^+$.
The situation changes at moderate shear rates, $\dot{\gamma} \in
[0.001; \; 0.005]\tau^{-1}$, where $J_s$ saturates and reaches a
maximum for all the isotherms. With the further increase of shear
rate, $\dot{\gamma} > 0.005\tau^{-1}$, the nucleation rate starts
to decrease. The rise of $n_c$ at shear rates $\dot{\gamma} \geq
0.001\tau^{-1}$ is due to the transformation of the nuclei from a
spherical form into prolate ellipsoids. This increase in size  and
the change in shape of the critical cluster at high shear rates
are directly reflected in the thermodynamic aspect of nucleation,
since the nucleation barrier is defined by the surface area and
the volume of the critical cluster. The appearance of maxima on
the curves $J_s(\dot{\gamma})$ illustrates the  antagonist  impact
of the shear flow  on the nucleation process: a  slow shear-flow
accelerates the nucleation trough the attachment rate, whereas the
high shear rates appear to destabilize the critical nuclei and
reduce the probability of the particle attachment. The physical
origin of the latter effect is discussed below. The fastest
nucleation rates are obtained for high temperatures and
intermediate shear rates, and are typically one order of magnitude
higher than in the absence of shear.  Note that the corresponding
shear rates would be achievable in colloidal suspensions under the
isothermal conditions considered here; in hard materials such as
metallic glasses, they would induce an important heating that is
not described by our calculations.

We mention  that the observed behavior of $J_s(\dot{\gamma})$ is
similar to the one detected for the same system confined by rigid
walls, undergoing inhomogeneous
flow~\cite{Mokshin/Barrat_2009_2010} and for     a   $2$D Ising
model under shear~\cite{Valeriani_JCP_2008}. Interestingly, the
nucleation rate $J_s$ takes lower values in the case of
homogeneous shear flow applied to a bulk glass (at the pressure
$P=14\epsilon/k_B$) in comparison  with results for a confined
system (at the pressure $P_{yy}=7.62\epsilon/k_B$) under
inhomogeneous shear drive~\cite{Mokshin/Barrat_2009_2010}.
Moreover, the maximum in $J_s(\dot{\gamma})$ observed in
Fig.~\ref{fig_nucl_rate} is shifted to higher  shear rates in the
case of an inhomogeneous shear. Nevertheless, the results of this
study provide a clear evidence that both types of shear flow --
homogeneous shear in a bulk glass and inhomogeneous shear in a
confined wall-driven glassy system -- yield the same homogeneous
nucleation mechanism. Moreover, the increase of
$J_s(\dot{\gamma})$ at low shear rates is similar with that was
observed by Graham and Olmsted in
polymers~\cite{Olmsted_PRL_103_2009}, albeit no saturation in
$J_s(\dot{\gamma})$ was detected in
Ref.~\cite{Olmsted_PRL_103_2009}. A simple extension the CNT to
describe the nucleation of colloidal suspensions under shear was
proposed in Ref.~\cite{Blaak_PRL_93_2004}, where the term $\Delta
G_{n_c}/k_B T$ and the critical cluster size $n_c$ were considered
to be  quadratic functions of the shear rate $\dot{\gamma}$. Such
a dependence does not seem consistent with the results for a
glassy system obtained above: although the cluster size $n_c$
grows with $\dot{\gamma}$ (see inset of Fig.~\ref{fig_nucl_rate}),
the dependence $n_c(\dot{\gamma})$ is not parabolic;, whereas the
dimensionless nucleation barrier $\Delta G_{n_c}/k_B T$ appears to
be a decreasing function of $\dot{\gamma}$ (see
Fig.~\ref{fig_eff_temp}b).

\subsection{Effective temperature}
The evolution of $T_{\mathrm{eff}}$ evaluated from Eqs.~(\ref{eq:
Zeldovich_factor_eff}) and (\ref{eq: surf_term}) is presented as a
function of $\dot{\gamma}$ in Fig.~\ref{fig_eff_temp}(a). The
extracted effective temperature reduces correctly to that of the
thermal bath in the limit of low shear rate,
$T_{\mathrm{eff}}(\dot{\gamma}=0) \rightarrow T$. This shows that
the CNT is perfectly consistent with our data in the shear-free
case. Interestingly, the systems quenched at low temperature are
undergoing ageing, which is a nonequilibrium process. Still, the
fact that the obtained temperature is identical to the one of the
thermal bath indicates that the fluctuations which trigger a
nucleation event are ``fast'', and they do not involve a slow
evolution within the free energy landscape. For shear rates
$\dot{\gamma} \geq 0.001\;\tau^{-1}$, a rise in
$T_{\mathrm{eff}}(\dot{\gamma})$ is observed, that is very similar
to the one detected in different studies of effective temperature in
sheared systems~\cite{Haxton_Xu_eff_temp}, where the effective
temperature was defined from fluctuation-dissipation relations. This
rise actually compensates the free energy cost of the nucleus
formation so that the Arrhenius factor $\textrm{exp}(-\Delta
G_{n_c}/k_B T_{\mathrm{eff}})$ increases with shear rate [see
Fig.~\ref{fig_eff_temp}(b)]. The inset in
Fig.~\ref{fig_eff_temp}(b), on the other hand, shows that the
non-monotonous behavior of $J_s$ arises from the
$\dot{\gamma}$-dependence of the kinetic prefactor, $g^{+}_{n_c}$.

Homogeneous shear flow results in  an anisotropy of nuclei growth, and
$g^{+}_{n_c}$ becomes direction dependent. At low temperatures,
the attachment rate averaged over directions depends on the strain
rate that initiates the motion, and on the probability that a
particle will remain attached to the
nucleus~\cite{Frenkel_book_1946}. It should, therefore, be
proportional to $\dot{\gamma}$ at small strain rates; while for
large strain rates, compared to the time scale $\tau_{\alpha}$ of
inherent structural rearrangements, the new configuration can be
destabilized before the attachment is achieved. The empirical form
suggested by the fits in figure \ref{fig_nucl_rate}, $g_{n_c}^+
\propto \dot{\gamma}\exp[- \xi(\dot{\gamma}\tau_{\alpha})^n]$,
accounts well for these different trends (here, $\xi$ and $n$ are
dimensionless parameters). In general, it can be expected that the
position of the  maximum in $g^+_{n_c}$  will correspond to
$\dot{\gamma_c}\propto 1/ \tau_{\alpha}$.

\section{Conclusion \label{conclusion}}
It appears quite remarkable, that a simple extension of classical
nucleation theory involving two independently determined quantities,
an effective temperature and a kinetic prefactor, describes
quantitatively the data for all the considered shear rates and
temperatures. Our interpretation of shear-driven ordering is
different from previously proposed approaches
\cite{Blaak_PRL_93_2004}, in which an empirical modification of the
nucleus free energy due to strain rate was suggested, while the
value of the temperature was kept fixed.  Here, we take the
alternative view that the shear flow produces additional
fluctuations, which enhance activated processes and can be described
by an effective temperature \cite{Ilg/Barrat}. As our analysis
allows one to completely disentangle the thermodynamic and kinetic
factors in the nucleation rate, we are moreover able to show that
the latter actually dominates the behavior of the nucleation rate at
high shear rates. Low shear rates promote nuclei growth by
increasing the mobility, but increasing the strain rate beyond a
critical, temperature-independent value results in a strong decrease
of the sticking coefficient.

\section{Acknowledgments} A.V.M. acknowledges  useful discussions
with R.M. Khusnutdinoff, A. Tanguy  and  E.D. Zanotto. J.L.B. is
supported by Institut Universitaire de France.

\bibliographystyle{unsrt}

\end{document}